\documentclass[useAMS,usenatbib]{mn2e} \usepackage{graphicx}
\usepackage{psfig} \usepackage{epsf}
\title[Neutron star binaries and long duration gamma-ray bursts]
{Neutron star binaries and long duration gamma-ray bursts}
\author[A.J.~Levan, M.B.~Davies and A.R.~King]{Andrew
J.~Levan$^{1,2}$\thanks{E-mail: anl@star.le.ac.uk (AJL);
mbd@astro.lu.se (MBD); ark@astro.le.ac.uk (ARK)}, Melvyn
B.~Davies$^{2}$ and  Andrew R.~King$^3$\\ $^{1}$Centre for
Astrophysics Research, University of Hertfordshire, College Lane,
Hatfield AL10 9AB, UK.\\ $^{2}$Lund Observatory, Box 43, SE--221 00,
Lund, Sweden.\\ $^{3}$Department of Physics and Astronomy, University
of Leicester, Leicester, LE1~7RH, UK.}
\begin{document}

\date{Accepted 2006 August 15 . Received 2006 August 5; in original
 form 2006 June 22}

\pagerange{\pageref{firstpage}--\pageref{lastpage}} \pubyear{2002}

\maketitle

\label{firstpage}

\begin{abstract}
Cosmological long-duration gamma-ray bursts (LGRBs) are thought to
originate from the core collapse to black holes of stripped massive
stars. Those with sufficient rotation form a
centrifugally-supported torus whose collapse powers the GRB. We
investigate the role of tidal locking within a tight binary as a
source of the necessary angular momentum. We find that the binary
orbit must be no wider than a few solar radii for a
torus to form upon core collapse. Comparing this criterion to the
observed population of binaries containing two compact objects
suggests that rotation may have been important in the formation of up
to 50\% of the observed systems. As these systems created a neutron
star and not a black hole they presumably did not produce highly
luminous GRBs. We suggest 
instead that they make
the subset of GRBs in
the relatively local universe which have much lower luminosity.
\end{abstract}

\section{Introduction}
The link between long-duration gamma-ray bursts (GRBs) and stellar
collapse is now firmly established (Woosley 1993; Hjorth et al. 2003;  Stanek et
al. 2003).  In particular long-duration GRBs appear to originate in type Ic
supernovae and frequently in hypernovae (core collapse events with an
order of magnitude more energy than classical supernovae).  The nature
of the short duration ($<2s$) bursts is far more uncertain,  although
their origin in populations of all ages (e.g. Gehrels et al. 2005;
Berger et al. 2005; Prochaska et al. 2006) can be explained if they
are caused by the final merger of a tight binary system of neutron
stars (NS) or black holes (e.g. NS-NS, NS-BH). In addition to these
two main classes of bursts, which are distinguished primarily based on
their observed duration, there is also evidence for further subtypes. For
example as  well as the very energetic GRBs which originate from high
redshifts  (e.g. a mean redshift of $\sim 2.8$; Jakobsson et al. 2006)
there is also  a further population of low luminosity events which can
be seen  only in the relatively local universe. The prototype for this
class is GRB~980425/SN~1998bw (Galama et al. 1998) which occurred only
35 Mpc away, while GRB~031203/SN~2003lw (at $\sim$ 450 Mpc; 
Watson et al. 2004; Malesani
et al. 2004) and most recently  GRB~060218/SN~2006aj (at $\sim$ 130
Mpc; Modjaz et al. 2006; Pian et al. 2006) also lie in this class.  
These low luminosity
events have isotropic equivalent luminosities  of only $10^{48 - 49}$ ergs
compared with energies of up to $10^{54}$ ergs for the most luminous GRBs.
Thus, while  the number of
observed systems is significantly lower, their space density is likely
to be much higher than for the more luminous bursts.

Regardless of the object responsible for the GRB (e.g. NS-NS binary or
collapsing star) the most popular model for the creation of the burst
is essentially the same - extreme accretion rates on to a
newly--formed compact object. This accretion is thought to be fuelled
by a massive (0.1-10 M$_{\odot}$) torus which forms if the infalling
material has too much angular momentum to accrete directly on to the
central compact object. In NS-NS mergers, it is relatively common for
mergers to produce a torus (see e.g. Ruffert \& Janka 1999;Rosswog \& Davies 2002).  For massive single stars, it is not clear
whether sufficiently high central rotation rates may be maintained to
produce a torus on core collapse (e.g. MacFadyen \& Woosley 1999;Petrovic et al. 2005), although
it has been suggested that rapidly rotating metal--poor stars can
retain sufficient angular momentum (Yoon \& Langer
2005; Woosley \& Heger 2006). Alternatively, binary scenarios suggest a way of removing the
envelope and providing a source of angular momentum (e.g. Izzard et
al. 2004; Podsiadlowski et al. 2004). We explore this mechanism
further in this paper.

Specifically we examine the idea that the core of a massive star may
be spun up by tidal locking within a tight binary.  The necessary
binary separations are a few solar radii or less.  The hydrogen
envelope of the massive star would therefore have been lost earlier
(probably in a common envelope phase). Systems sufficiently tight may 
undergo a further period of mass transfer, similar to that seen in Cyg-X2 
(e.g. Davies, Ritter \& King 2002).
We subsequently examine known
compact binary systems (NS-NS, NS-WD) and conclude that, at the time
of the supernova, approximately 50\% may have been sufficiently 
tight to necessitate the formation of a disc.  

We suggest that low and high luminosity GRBs may be distinguished by
the formation of either a NS (low luminosity) or BH (high
luminosity). The only compact object -- compact object
binaries discovered so far in the Milky Way consist of
neutron stars and white dwarfs. 
However similar systems, with more
massive cores could produce black holes, and thus luminous GRBs.

\section{Evolutionary pathways to compact object binaries} 
\begin{figure}
\begin{center}
\resizebox{6truecm}{!}{\includegraphics{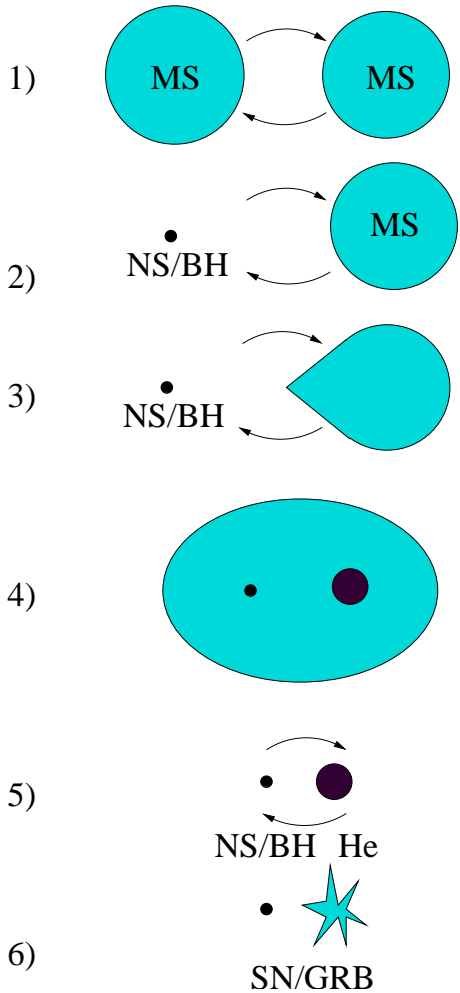}}
\end{center}
\caption{An evolutionary pathway to the creation of a
binary containing a rapidly-rotating core-collapse supernova 
in a tight orbit.
The primary evolves first, possibly transferring material to
the secondary (stage 1). It then produces a neutron star (NS) 
or black hole (BH),
when it explodes as a core-collapse supernova (stage 2).
The secondary then evolves, filling its Roche lobe (stage 3) and
transferring material to the NS/BH producing a common envelope 
phase (stage 4). The NS/BH and He core of the secondary spiral
together ejecting the surrounding envelope producing a very 
compact binary (stage 5). Tidal locking produces a rapidly-rotating
He star such that the rotation is significant when the secondary
explodes as a core-collapse supernova, with a torus being formed
around the central compact object by infalling material.} 
\label{chan}
\end{figure}

Compact binaries consisting of some combination of NS and BH, and
possibly white dwarfs (WDs), can be formed via a variety of channels
(see e.g. Belczynski et al. 2002). The basic scheme is shown in
Figure~\ref{chan}, and is essentially the same as that described in
Bhattacharya \& van den Heuvel (1991).

The route to forming the compact object binary is thus a relatively
close binary system. For the formation of a double neutron star
(rather than a NS-WD system) both components must have $M >
8$M$_{\odot}$.  The initially more massive star evolves more rapidly,
leaving the main sequence before its companion and, if the binary is
sufficiently close, causing a first incidence of conservative mass
transfer. This phase of mass transfer may significantly increase the
mass of the secondary (as was the case for J1141-6545 [Davies et
al. 2002]).  The core of the primary continues to evolve to the
point of core collapse and supernova explosion, producing either a
neutron star or black hole depending on its mass. Subsequently the
second star evolves off the main sequence. As the mass ratio at this
point is large, the resulting runaway mass transfer creates a common
envelope in which the He core of the secondary and the first neutron
star inspiral due to the loss of orbital angular momentum and energy
via dynamical friction in the envelope.  Eventually this envelope is
removed and a He-star -- NS system remains (i.e.  the hydrogen has now
been removed from the system so the final SN is Type I).

The penultimate step in the evolution of compact binaries is the
evolution of the He star -- NS binary. As the final
process prior to the second SN, the evolution of the binary at this
stage may have the largest impact on dynamics of the SN upon
collapse. In particular the angular momentum of the He-star
largely dictates the formation (or not) of a disc upon core
collapse. We assume here that the binary is tidally locked at the
end of the helium main sequence, this is reasonable since
the timescale for synchronisation of the orbit is less than the
evolutionary timescale for the Helium star (Hut 1981;Tassoul 1995).
Given this, we then calcluate 
the angular momentum content of the core at the time of core collapse, as
discussed in the next section.  The evolution of a He-star -- NS
binary has been investigated in detail by  Dewi et al. (2002, 2003). For the
very close binaries of interest here the He star essentially
always fills its Roche lobe, either on the helium  main sequence or in
the giant branch. We use a simplistic model in which the
subsequent evolution of the orbit is governed by
three factors; i) inspiral via gravitational radiation, ii) 
mass transfer and iii) mass loss via winds. Gravitational
radiation always acts to decrease the separation of the two components,
while mass loss decreases the total mass of the system and may drive the
binary to wider separations. We do not consider the effect of
magnetic fields within the star, although they may act to slow the rotation
(e.g. Petrovic et al. 2005). 
This may occur since the core and envelope remain magnetically
coupled after the He main sequence, or because of 
angular momentum loss via stellar winds flowing along
the field lines while on the main sequence. The latter effect would
be less noticeable at lower metallicity and may explain 
a bias towards GRB in low metallicity environments (see section 7).
The effect of mass transfer when the He
star overflows its Roche lobe can be more complex. Matter overflowing
the Roche lobe falls onto the already formed neutron star and angular
momentum may be ejected from the binary (e.g. via a wind).
However, if this
material  has low specific angular momentum then the system can
widen. Indeed, Dewi et al. (2002) find that for massive He stars
(i.e. those that form neutron stars) the period typically
increases through the He star evolution.  
Alternatively, mass transfer from the He star may result in the merger of the
He star and the neutron star via a delayed dynamical
instability if the mass ratio is too extreme.

\section{Discs around neutron stars and black holes}
In the general model discussed above, rotation is a crucial
ingredient in the  production of a GRB. Too little rotation 
results in the direct collapse of the progenitor star and will preclude
the formation of a torus and thus of a GRB.  Here we assume that  the
material lying just outside the core must be centrifugally supported
upon collapse, in order to form an
accretion disk outside the innermost  stable orbit of the nascent
black hole.
This criterion can be expressed quantitatively as requiring that
the specific angular momentum $j$ of the material just outside the
core exceeds
$\sqrt{D} G M_{\rm c} / c$, where  $D$ is the
radius of the orbit required for disk formation in units of
Schwarzschild radii ($GM/c^2$)
(e.g. the innermost stable orbit lies at $D=6$, or
$\sim$ 12 km for an 1.4 $M_{\odot}$ BH), $M_{\rm c}$ and the core mass 
(Podsiadlowski et al. 2004).  

Discs around neutron stars can be more difficult to create, since
young neutron stars can be markedly larger than the canonical 10 km
radius commonly assumed. Therefore the disk must be formed at higher
($\sim 20-50$ km) radius from the newly formed NS.  The creation of
the disk at a larger radius thus requires greater angular momentum.
If this requirement arises from a tidally locked binary, we can show
that this in turn demands
a correspondingly higher orbital velocity.
We assume as described above that tidal locking occurs at the
beginning of the Helium main sequence and that, for the late stages of
evolution of the He-star (i.e. in the He giant branch) the core
decouples from the envelope.  In this case, following Podsiadlowski et
al. (2004) we can compare the required angular momentum at the time of
collapse with that at the edge of the iron core at the start of the
helium main sequence, and, assuming tidal locking equate this to an
orbital frequency (i.e.  $\omega = \sqrt{D} G M_{\rm c} / R_{\rm c}^2 c$, where
$R_{\rm c}$ is the radius of the iron core). Assuming synchronous rotation,
this gives a critical orbital separation

\begin{equation} 
a < (4M_{\rm tot} c^2 R_{\rm c}^4 / 9DG M_{\rm c}^2)^{1/3},
\end{equation}
or
\begin{equation} 
\left( {a \over R_{\odot}} \right) < {60 \over D^{1/3}} \left({R_{\rm c}
\over R_{\odot}}\right)^{4/3} \left({M_{\rm tot} \over
M_{\odot}}\right)^{1/3} \left({M_{\rm c} \over M_{\odot}}\right)^{-2/3},
\end{equation}
where $M_{\rm c}$ and $R_{\rm c}$ are the core mass and radius, and
$M_{\rm tot}$ is the total mass of the binary.  We use here
the models of Heger et al. (2002), the key parameters are
$M_{\rm c} = 1.7$ M$_{\odot}$, $R_{\rm c}
\sim 0.1 R_{\odot}$, for a He star of mass 7.71 M$_{\odot}$, this
yields a critical separation of $\sim 3$ R$_{\odot}$ for the formation
of a disc at the innermost stable orbit of a 1.7 $M_{\odot}$ BH.

\section{Comparison with known systems}

\begin{figure}
\begin{center}
\resizebox{8truecm}{!}{\includegraphics{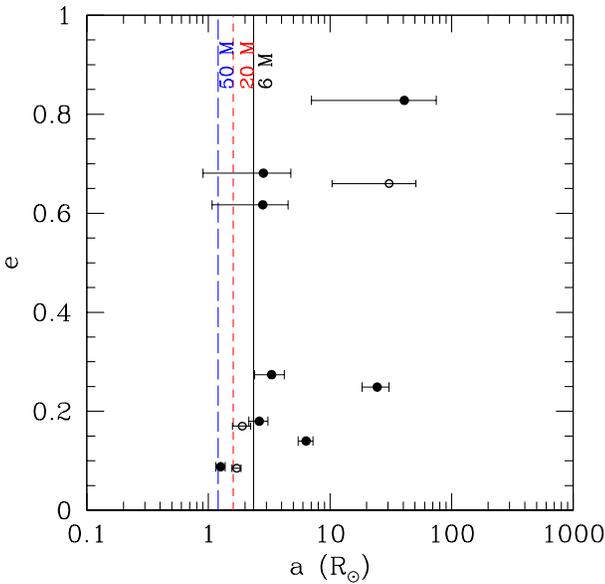}}
\end{center}
\caption{The eccentricity and semi-major axis of observed systems
containing either two neutron stars, or a neutron star and white
dwarf.  The error bars give an indication of the range of separations
between the two stars during their orbit. Neglecting the inspiral the
separation at the time of the supernova must be taken from this range
(see Table 1 for the calculations including the effects of
inspiral via gravitational radiation).  
The three vertical lines represent the critical separations necessary
at the time of core collapse for a centrifugally supported disc to have
formed at a distance of 6M, 20M and 50M from the newly formed 
compact object,  based on the models of
Heger et al. (2002) described in section 3 (where $M= GM_{ns/bh}/c^2$ and is 
$\sim 12, 40$\& 100 km for D=6,20 \& 50 for a 1.4 M$_{\odot}$ NS). Lower
total masses, as were likely the case of J0737-3039 require
slightly tighter orbits, although the orbit is only a weak function
of the total mass ($M_{tot}^{1/3}$).
 If the binary separation is less than this (i.e. 
to the left of the line) then disc formation is favoured.
The data are from Champion et al. (2004)
and Lorimer (2005). In cases where the masses of the two components
have not be measured 1.4 M$_{\odot}$ has been assumed. Known NS-NS
binaries are indicated with circles, while NS-WD or those with
uncertain companions are marked with open circles.
}
\label{ae}
\end{figure}

We have examined the distribution of separations for the {\it known}
compact binary systems (NS-NS or NS-WD) within our own Galaxy.  These
are shown in the semi-major axis -- eccentricity plane in Figure \ref{ae}.
From the currently--measured orbital parameters it is possible to constrain
the pre-explosion parameters of each of the systems. The difference
between the currently observed orbital parameters, and those
immediately prior to the SN comes from the inspiral of the binary components
via gravitational radiation  and the effects of the kick imparted to the
second NS at the time of the SN.  Both of these effects
can be accounted for. Initially the  inspiral of the orbit can be
extrapolated back to the time of the second SN using the
equations of Peters (1964) for the time evolution of eccentricity
and semimajor axis for a binary. 
Secondly the range of pre-SN separations can be
estimated, since, as the orbit is closed, the NS must return to the
point in the orbit at which the  SN occurred, i.e.
\begin{equation}
(1-e_{SN})a < a_{SN} < (1+e_{SN})a,
\end{equation}
where $a_{SN}$ and $e_{SN}$ represent the semi-major axis and
eccentricity of the system, immediately after the second SN, as
calculated by integrating the Peters (1964) relations back to the
characteristic age of the recycled pulsar.  The results of this are
shown in Table 1, and demonstrate that only modest evolution of the
orbit has occurred.

As can be seen in Figure \ref{ae} a number of compact binaries have
separations which are sufficiently close that, upon collapse material
may have been rotating too rapidly to fall directly onto the nascent
NS and thus it  is expected that a disc will have  formed. This represents
roughly 50\% of the observed population of compact binaries, and the
relevant systems are tabulated in Table 1.

However, there are reasons to suspect that this measure of the
fraction of  compact binaries forming discs may be an
underestimate. For example population synthesis models
(e.g. Belczynski et al. 2002) show that a large fraction of double
neutron star binaries form with orbit separations $\ll 1 R_{\odot}$ and
subsequently have merger times of $<10^6$ years.  This population is
under-represented in current pulsar surveys since their short
lifetimes bias against their discovery even though (by number formed)
they may be the dominant population. Such NS-NS binaries {\it must} at
the time of formation of the second NS have had separations small
enough to  meet the criteria described above, and it is thus
reasonable to suspect that disc formation should have occurred in each
of these systems at the time of core collapse.

\begin{table*}
\begin{center}
\begin{tabular}{llll|llll}
\hline
System						& J0737-3039 	& J1906+0746 & J1141-6545 & J1756-2251 & B2127+11C & B1913+16 & B1534+12\\
\hline
$M_{total}$ (M$_{\odot}$) 		&2.58	&2.61	&2.31  & 2.57 & 2.71 & 2.83 & 2.75\\
$P_{orb}$ (days)				&0.102	&0.16	&0.20  & 0.32 & 0.3   & 0.3  & 0.421\\
$a_{orb}$ ($R_{\odot}$) 			&1.3		&1.7 	&1.9     & 2.7 & 2.8     & 2.8 & 3.3\\
$e$							&0.088	&0.085 	&0.17   & 0.18 & 0.68 &  0.62 & 0.274\\
age ($10^8$) years				& 0.5	&0.001 	&0.014 & 4.4 & 0.97   & 1.1 & 2.5\\
$a_{SN}$ ($R_{\odot}$)			&1.45	&same 	&same & 2.9 & 3.51    & 3.25 & 3.39\\
 $e_{SN}$ 					&0.10	& same 	&same & 0.20 & 0.74  & 0.66 & 0.282\\
 $a_{min},a_{max}$ ($R_{\odot}$)	&1.3,1.6	&1.6,1.8 	&1.6,2.2 & 2.2,3.2 & 0.9,6.1 & 1.1,5.4 & 2.4,4.2\\
\hline
\end{tabular} 
\end{center} 
\caption{Properties of compact binaries which may have formed a disc
during the second SN explosion.  The current orbital parameters are
from Champion et al. (2004), Lorimer (2005) and Faulkner et al. (2005) while
the $a$ and $e$ immediately after the second SN have been calculated
by extrapolating the evolution of the orbit via gravitational
radiation (Peters 1964) back over the characteristic age of the
pulsar. $a_{min}$ and $a_{max}$ represent the range of plausible
separations of the binary at the time of the second SN and can be
compared to the critical values for the formation of a disc at
different radii as shown in Figure~\ref{ae}}
\label{default}
\end{table*}%

\begin{table}
\begin{center}
\begin{tabular}{lll}
\hline & Galaxy$^{-1}$ yr$^{-1}$ &	Gpc$^{-3}$ yr$^{-1}$\\ \hline
	GRBs &               $1 \times 10^{-7}$	&	1.5 \\ 
	LLGRBs &          $3\times 10^{-5}$&	500 	\\ 
	SGRBs &	           $>2 \times 10^{-6}$&	$>$30\\ 
	SN CC &             $7 \times 10^{-3}$& 100000 \\
	HNe &                 $ 1 \times 10^{-5}$	&$150$\\
	SN Ic & 	          $1 \times 10^{-3}$	& 15000\\ 
	NS-NS &		 $1 \times 10^{-4}$ & $1500$\\ 
	NS-BH &		 $1 \times 10^{-5}$ & $150$ \\ 
	BH-BH &             $3 \times 10^{-5}$ & $450$\\ 
	\hline
\end{tabular}
\end{center}
\caption{Approximate local rates of different events related to GRBs
including  Low Luminosity GRBs (LLGRBs - Liang et al. 2006) and short
GRBs (SGRBs - Nakar et al. 2005). The rates have been taken (or
derived) from Podsiadlowski et al. (2004), Kalogera et al. (2004),
Belczynski et al. (2002) and Cappellaro et al. (1999) }
\label{default}
\end{table}%

\section{NS-NS binary formation and low luminosity GRBs}
Figure~\ref{ae} clearly shows that, at the time of core collapse,
several  of the observed NS-NS binaries were sufficiently close for
centrifugally  supported discs to form. Since GRBs are commonly
thought to  originate from similar discs surrounding black holes, it is
interesting to  investigate whether SN producing discs around newly-formed
neutron stars may also produce some form of GRB.
The maximum energy released in the
accretion  of the torus is given by $E = G M_{\rm ns/bh} M_{\rm acc} /
R_{\rm ns/bh}$,  or $E_{\rm ns} = 3.6 \times 10^{53} (M_{\rm acc} / M_{\odot})$
ergs, for a 1.4 M$_{\odot}$ NS and $E_{\rm bh} = 1 \times 10^{54} (M_{\rm acc}
/ M_{\odot})$ ergs, for any mass BH (since $M_{\rm bh} \propto R_{\rm bh}$).

The extrapolation from disc accretion to gamma-ray luminosity is far
from trivial since it requires an assumption about the conversion of
accretion luminosity into $\gamma$-ray energy. One plausible mechanism
of providing this energy is via neutrino-antineutrino annihilation.
If this is assumed to be the energy source then the accretion energy  can
be related to the observed gamma-ray energy via several efficiency
factors which account for the conversion of accretion energy to
neutrinos, the cross section for neutrino - antineutrino annihilation,
the subsequent fraction of energy that is transferred into a baryon
free jet, and finally the fraction of this energy which is emitted as
gamma-rays (Oechslin \& Janka 2006).  Following Oechslin \& Janka
(2006) we assume that the product of the these efficiencies is $\sim
10^{-3}$. Thus the observed luminosities of  low luminosity GRBs of
$10^{48} - 10^{50}$ ergs can be  explained by the accretion of $ 0.01
< (M_{acc} / M_{\odot}) < 0.3$ of material from the disc.

There is, of course, a fundamental limit to the mass which can be
accreted onto a NS without the consequent accretion induced collapse
(AIC) of the neutron star on to a BH (e.g. Dermer \& Atoyan 2006). Roughly speaking if the NS mass
exceeds $\sim 3$ M$_{\odot}$  then a BH forms. The masses of
the neutron stars seen in NS-NS binaries are typically close to 1.4
M$_{\odot}$ ($q \sim 1$), so
 accretion from
the disc has not added dramatically to the mass.
We therefore 
suggest that the result may have been a GRB with a low luminosity of
$10^{48} - 10^{50}$ ergs. These systems may therefore be the
progenitors of the nearby population of long duration GRBs with lower
energies but significantly higher space densities.

As described above, the final step in the formation of a NS-NS binary
is the collapse of either a helium or carbon-oxygen core of a massive
star in a type Ib/c supernova. Such supernovae only represent a
fraction 15\% of the total core collapse SN population (Podsiadlowski
et al. 2004), and those in tight binary systems only a small fraction
of these -- approximately 3-10 \%  based on the ratio of SN Ib/c to
compact object -- compact object binaries formed (Belcyznski et
al. 2002)\footnote{Note:  this is formally a lower limit on the ratio
of binary to single SN Ib/c, since a fraction of NS-NS binaries
are disrupted on formation of the second NS. This fraction is likely to
be small for very tight binaries but larger for wider systems.}
 However, it is worth noting that the pathway to NS-NS
production described in this paper and the creation of a long-duration
GRB both require an SN Ib/c event (e.g. Hjorth et al. 2003; 
Pian et al. 2006).

From observations of the (relatively small) number of low
luminosity GRBs observed to date their local space density is
estimated as 700$^{+1400}_{-500}$ Gpc$^{-3}$ yr$^{-1}$(Soderberg et
al. 2006) and 522 Gpc$^{-3}$ yr$^{-1}$(Liang, Zhang \& Dai 2006). This compares
with a rate a local rate of NS-NS mergers which lies in the range
200-3000 Gpc$^{-3}$  yr$^{-1}$
(Kalogera et al. 2004; Nakar et al. 2006),
although significant uncertainties  remain in the derivation of each
of these rates it is likely that there are sufficient  compact
binaries (NS-NS, NS-WD, BH-NS, BH-BH) systems to explain  the
population of low luminosity GRBs.

\section{High Mass Analogues and Cosmological Long GRBs}
The observed compact object -- compact object binaries in the 
Milky Way consist either of WDs or
NSs. 
No BH-NS or BH-WD or BH-BH systems are known. However, it is
expected that they are formed in moderate numbers; they are
simply harder to detect compared to NS-NS systems
which have been found via the radio emission from
the recycled pulsars.  BH-BH systems can be formed
via the same channel as that shown in Figure~\ref{chan} but with
initially higher main sequence masses.  In the common model for the
creation of GRBs a nascent BH at the core of the massive star is
essential. Thus, more massive versions of the systems which formed the
observed NS-NS binaries are candidates  for the formation of classical
long-duration GRBs.

In this context it is reasonable to examine the parameter space which
may allow the formation of a rapidly rotating black hole within
tidally locked binary systems.  In particular, for two
massive
stars, both of which create a BH 
we wish to know
whether both of the SN could
plausibly produce a GRB or if the presence of a compact object prior
to the GRB forming SN aids the production of GRBs.

Figure 3 shows constraints
on the radius of the secondary object such  that it remains within its
Roche lobe at the critical orbital separation. If both the He star and
the companion overflow their Roche lobes then a further common
envelope results, and the most likely scenario is a merger
of the He star and the companion. For this
constraint to be met the radius of the secondary must be $\ga$ 1
R$_{\odot}$. Thus only low-mass main-sequence stars could
remain within their Roche lobes in these binaries.  However, should
the companion be a compact object (BH or NS) at the time when the
progenitor He star enters the He main sequence then the evolution will
proceed as described in section 2. In this scenario it is worth
noting that the creation of a BH-BH system can aid the production
of a GRB in two further ways. Initially
the critical rotation periods can be attained 
at larger separations
for  more massive companions. In addition, mass transfer
from the star is less likely to lead to a delayed dynamical 
instability when the secondary is a black hole. This is because
the mass ratio is lower with a more massive black hole companion
rather than a neutron star.

In some cases, when either the mass ratio is too extreme, or the 
He -- NS/BH separation is too small following the common envelope, the
binary may merge (e.g Fryer \& Woosley 1998),
such cases can also plausibly create
GRBs by either accretion onto the BH in He -- BH systems (Fryer \& Woosley 1998),
or by merger of two He stars if the masses are sufficiently similar (Fryer \& Heger 2005;
Dewi et al. 2006). Given that under our criteria a reasonable
fraction of the observed systems would have formed a disc 
it appears unlikely that all binaries with sufficient angular momentum
for GRB production undergo such mergers.

\begin{figure}
\begin{center}
\resizebox{8truecm}{!}{\includegraphics{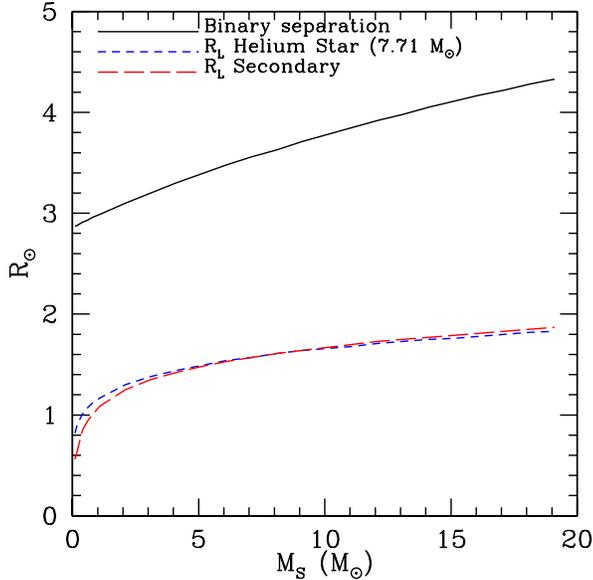}}
\end{center}
\caption{The maximum separation allowed for binaries containing a
helium core of a massive star, as a function of secondary mass, such
that core collapse of the helium star leads to the formation of a
torus of material around the central core, assuming that the system is
tidally locked.  The Roche lobe radii for the two stars are also
shown. As can be seen, main sequence stars with radii $>$ 1$R_{\odot}$
cannot  fit within their Roche lobes, and, assuming $M \propto R$
this sets a mass limit of
$\sim$ 1 M$_{\odot}$ on the mass of a main sequence companion. 
However, should
both stars overflow their Roche lobes a common envelope would
follow and, most likely the resulting system would merge.}
\label{rl}
\end{figure}

\section{Discussion}
Several lines of recent evidence have pointed to a preference 
by GRBs for 
host galaxies of low metallicity (Fynbo et al. 2003; Fruchter
et al. 2006;  Stanek et al. 2006). In contrast the systems 
observed in the Milky Way presumably formed at higher
metallicity (probably $\sim Z_{\odot}$). However, one may expect 
that similar systems form GRBs more readily in
lower metallicity environments,
due largely to the effects of mass loss on the evolution
of the binary. Mass loss from both stars on the main
sequence will drive the binary to wider orbits, reducing
the fraction of binaries which come into contact at
any stage and also those which are sufficiently tight
for disc formation upon core collapse.  Additionally, mass
lost by the Helium star on its main sequence is lost along
the magnetic field lines and carries matter to large radii, losing 
angular momentum from the stellar core and braking its rotation.
At lower metallicities the low mass loss rates thus favour more
rapid core rotation.
Finally, mass loss from the individual stars
can dictate the remnant left (NS or BH). In lower metallicity
environments the lower mass loss rates thus leave the
binary in a closer orbit, and with more massive stellar cores,
which collapse to form BHs. Thus lower metallicity will favour
the production of GRBs, in agreement with observations.

One consequence of a supernova within a close binary is that the
outflowing supernova ejecta must pass the older NS. For close binaries
the density at the radii of the NS is moderately high and thus, even
though the velocity of the ejecta can be very high the Bondi--Hoyle
accretion rate may be significant (see e.g. Broderick 2005). 
Accretion on to this object may provide a additional source of energy,
and thus lightcurve variability, 
in any SN/GRB formed via this
channel.

It is interesting to note that 
in this model some of
these systems can
create both a long and a short GRB, with the short burst occurring
some time ($<10^8$ years) after the long burst. 
We note that
the NS-NS systems 
discussed here
are distinct from those found in globular
clusters, which have been suggested as short GRB progenitors (Grindlay
et al. 2005), since NS-NS systems in globular clusters are formed
dynamically and not by the channel  described here.

\section{Conclusions}
We have examined the role of tidal locking in a tight binary system
and its implications for the formation of a centrifugally-supported
 torus upon
core collapse. By examining the  distribution of separations of known
compact object binaries (NS-NS, NS-WD), we conclude that up to $\sim$
50\% of the systems could, at the time of the second  supernova, have
been sufficiently rapidly rotating to create a torus upon collapse.
We subsequently investigated the implications of these observations for
the  formation of GRBs. None of the observed systems contain a BH,
which is  thought to be essential to the formation of the highly
luminous GRBs. We suggest that higher mass analogues of the observed
systems can form GRBs, but that the collapse of less massive stars
leading to neutron stars might result in  a lower luminosity GRB,
similar to those that have been observed in the local  universe.

\section*{Acknowledgments}
We thank the referee for a very helpful report.
AJL is grateful to PPARC for a postdoctoral fellowship award. AJL also
thanks the Swedish Institute for support while visiting Lund.  MBD is
a Royal Swedish Academy Research Fellow supported by a grant from the
Knut and Alice Wallenberg Foundation.  ARK gratefully acknowledges a
Royal  Society--Wolfson Research Merit Award.

\label{lastpage}

\end{document}